\documentclass[]{raa}
\usepackage{graphicx,times}
\usepackage{natbib}
\usepackage{amsfonts}
\usepackage{txfonts}

\providecommand{\bm}[1]{\mbox{\boldmath$#1$\unboldmath}}

\begin{document}
\title{Contradiction between strong lensing statistics and a feedback solution to the cusp/core problem}

\volnopage{ {\bf 2010} Vol.\ {\bf XX} No. {\bf XX}, 000--000}
   \setcounter{page}{1}

\author{Da-Ming Chen\inst{1} 
\and Stacy McGaugh\inst{2}}
\institute{National Astronomical Observatories, Chinese Academy of Sciences,
Beijing 100012, China; {\it cdm@nao.cas.cn} \\
\and
Department of Astronomy, University of Maryland, College Park, MD
20742-2421, USA}

\vs \no
   {\small Received [2010] [July] [7]; accepted [2010] [Sep] [22] }

\abstract{
Standard cosmology has many successes on large scales, but faces some
fundamental difficulties on small, galactic scales. One such difficulty is the
cusp/core problem. High resolution observations of the rotation curves for dark
matter dominated low surface brightness (LSB) galaxies imply that galactic dark
matter halos have a density profile with a flat central core, whereas N-body
structure formation simulations predict a divergent (cuspy) density profile at
the center. It has been proposed that this problem can be resolved by stellar
feedback driving turbulent gas motion that erases the initial cusp. However,
strong gravitational lensing prefers a cuspy density profile for galactic halos.
In this paper, we use the most recent high resolution observations of the
rotation curves of LSB galaxies to fit the core size as a function of halo mass,
and compare the resultant lensing probability to the observational results for
the well defined combined sample of the Cosmic Lens All-Sky Survey (CLASS) and
Jodrell Bank/Very Large Array Astrometric Survey (JVAS). The lensing
probabilities based on such density profiles are too
low to match the observed lensing in CLASS/JVAS.  High baryon densities in
the galaxies that dominate the lensing statistics can reconcile this discrepancy,
but only if they steepen the mass profile rather than making it more shallow.
This places contradictory demands upon the effects of baryons on the central
mass profiles of galaxies.
}

\authorrunning{Chen \& McGaugh}
\titlerunning{Strong Lensing and Cusp/Core Problem}
\maketitle

In the standard cosmological model (known as $\Lambda$CDM), the
universe is dominated by invisible components called dark energy ($\Lambda$) and
cold dark matter (CDM). The  $\Lambda$CDM cosmology is very successful in
explaining the cosmic microwave background and the formation of large scale
structure. However, there are challenges to $\Lambda$CDM on smaller
scales \citep{coles05}. Here we focus on the cusp/core
problem \citep{nfw97,jing00,jing02,nfw04,li09} and whether proposed solutions to this problem
can be consistent with the observed frequency of strong gravitational lensing.

One possible solution to the cusp/core problem is turbulence driven by stellar
feedback during galaxy formation.  If this process drives massive clumps of gas
through the central regions of the first dark matter halos to
form \citep{mcw06,mcw08}, the central cusp may transform into a soft core.  Once
established, phase space arguments imply that the core should persist through
subsequent mergers \citep{dehnen05,kzk06}, leading to a final halo profile with a
finite core radius for all galaxies, including giant ellipticals.
Such a situation is consistent with
essentially all kinematic observations \citep{mcgaugh07,roman03}.
The stellar feedback model is claimed to be `universal' to all masses of galaxies, so it should be verified
by observations of galaxies not only with low mass like dwarfs and LSBs, but of all masses, especially
large mass galaxies like giant ellipticals.  We show here that if stellar feedback solution to the cusp/core problem (arising from low mass LSB galaxies) is true, then
it should also pass the tests of the observations of massive galaxies, in particular the observations of strong gravitational lensing. To do so, we extrapolate the core size-halo mass relation established from rotation curve data of low mass galaxies to massive ellipticals so that we can calculate the strong lensing probabilities.

\begin{table}
\caption{Halo Profiles}
{\tiny
\begin{center}
\begin{tabular}{lcc}
\hline\\
Halo & $\gamma$ & $\rho(r)$ \\
\hline\\
SIS & 2 & $\rho_0 (r/r_0)^{-2}$ \\
NFW & 1 & $\rho_i [(r/r_s)(1+r/r_s)^2]^{-1}$ \\
CIS & 0 & $\rho_0 [1+(r/r_c)^2]^{-1}$ \\
\hline\\
\end{tabular}
\end{center}
}
\end{table}

Gravitational lensing provides a powerful tool to detect dark matter. The
lensing efficiency is very sensitive to the slope $\gamma$ of the central mass
density profile ($\rho \propto r^{-\gamma}$). It is well
established \citep{chae02,li02,oguri08} that when galaxies are modeled as a
singular isothermal sphere (SIS: $\gamma = 2$) and galaxy clusters are modeled
as a Navarro-Frenk-White (NFW: $\gamma = 1$) profile (see Table 1), the
predicted strong lensing probabilities match the results from CLASS/JVAS. A
steeper density slope near the center gives a more efficient lensing rate. For
example, if we model galaxies with an NFW rather than SIS profile, the lensing
probabilities are too low compared with observations at small image
separations \citep{li02}. The presence of a central flat core ($\gamma \approx 0$)
in galaxies would further limit the lensing efficiency \citep{chen05}. For
example, a nonsingular truncated isothermal sphere (NTIS), which is an
analytical model \citep{shapiro99} for the postcollapse equilibrium structure of
virialized objects, has a soft core that matches quite well with the mass
profiles of dark matter dominated LSB galaxies deduced from their observed
rotation curves. The probabilities for lensing by NTIS halos are far too low
compared to observations \citep{chen05}, however.

In order to investigate the effect of a central core on the strong lensing
efficiency, we use the density profile of the halos directly constrained by
observed rotation curves. These are well fit \citep{begeman91} by the cored
isothermal sphere (CIS).  The CIS halo has a finite core radius $r_c$ within
which the density is constant ($\gamma = 0$).  As well as providing a good
description of the data, the CIS provides a reasonable proxy for the unwieldy
NTIS profile.
Initially, we consider lensing by the dark matter halo itself,
and later consider the additional effects of the baryons.

\begin{figure}
\includegraphics[width=8.5cm]{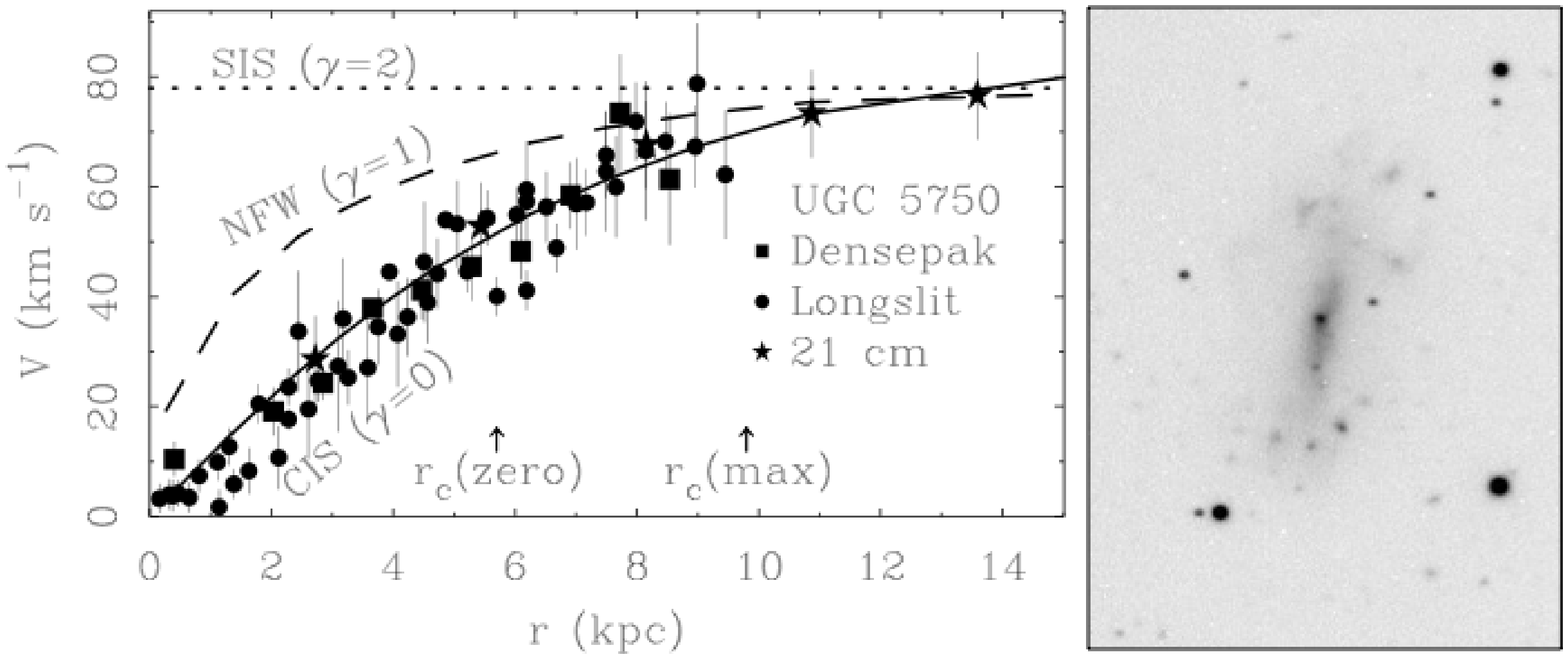}
\caption{The rotation curve (left) of the LSB galaxy UGC 5750 (right).
Velocity data come from several independent sources and methods, including
radio synthesis observations of the 21 cm spin flip transition of atomic
hydrogen \citep{vdHulst93}, two independent \citep{MRB01, deblok02}
optical long slit observations of the n=3$\rightarrow$2 Balmer transition
(H$\alpha$), and Densepak integrated field H$\alpha$
spectroscopy \citep{KdN06}.  The various halo types are shown as lines
(as marked).  The parameters of NFW halos are not free,
following \citep{nfw97,KdN06} from $\Lambda$CDM cosmology.
The difference between this and the data is the cusp/core
problem.  The core radius of the CIS fit is marked by arrows for the
cases of zero and maximum disk.  For clarity, the full CIS halo is only shown
for the case of zero disk.  Attributing mass to the stellar disk detracts from
the velocity that can be attributed to dark matter (albeit not much in the case
of LSB galaxies), increasing $r_c$ as shown and makes the discrepancy
with the NFW prediction of $\Lambda$CDM more serious.  Under no
circumstances can the halos of LSB galaxies be modeled by SIS.
\label{u5750RC}}
\end{figure}

The best objects for tracing the mass profile of the dominant dark matter
component are LSB galaxies.  In other galaxy types, the stellar mass can
provide a non-negligible contribution to the rotation velocity at
observed radii.  This is not the case for LSB galaxies, whose diffuse
disks remain dark matter dominated \citep{dBM97} down to small radii.
These objects persistently suggest that dark matter halos possess
approximately flat cores \citep{dBMBR01} that are best fit with CIS halos
(Figure \ref{u5750RC}).

We use the most recent results \citep{denaray08} from a sample of LSB galaxies for
which rotation curves have been derived from high-resolution optical velocity
fields. For each halo, we calculate the mass $M$ by integrating the CIS density
profile to the radius $r_{200}$. This is the radius of a sphere within which the
average mass density is 200 times the critical density of the universe,
typically taken \citep{nfw97} as the virial radius,
\begin{eqnarray}
M&=&M_{200}=\int_0^{r_{200}}4\pi\rho_{cis}(r)r^2dr  \nonumber \\
&=&4\pi\rho_0r_c^2[r_{200}-r_c\arctan(r_{200}/r_c)]
\label{mass}
\end{eqnarray}.
We compute the halo mass for two bracketing assumptions \citep{denaray08}
about the mass of the baryonic disk:  zero disk, in which the mass of stars and
gas is neglected, and maximum disk, which attributes the most mass possible to
the stars without exceeding the observed rotation.  The primary
difference between these two cases is in the core radius inferred for the halo.
As more mass is attributed to the stars, less dark matter is necessary at small
radii.  Consequently, $r_c$ grows with stellar mass.

\begin{figure}
\includegraphics[width=8.5cm]{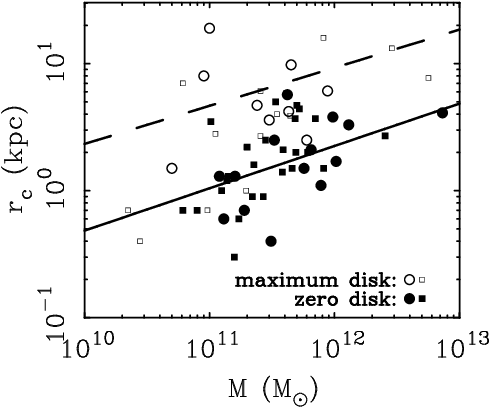}
\caption{The correlation between core radius $r_c$ and halo mass $M$. Filled
points represent the case of zero disk mass and the open points represent
maximum disk. The lines are the fit to the
Densepak data \citep{denaray08} only (circles); fits to long slit
data \citep{deblok02} (squares) give indistinguishable resuts.
\label{MRc}}
\end{figure}

There is an established correlation between $\rho_0$ and $r_c$ that can be
fitted with a power-law formula \citep{kormendy04}. Then together
with $M_{200}=(4\pi/3)r_{200}^3\times 200\times\rho_{crit}$, equation (\ref{mass}) can be numerically solved for
any $M$ and the solution can be approximated by a power-law formula \citep{salucci07}.
We do not fit $\rho_0$ and $r_c$, instead, for each halo of the sample,
we substitute the corresponding $\rho_0$ and $r_c$ into equation (1) to numerically obtain $M$,
then fit  $r_c$ and $M$ with a power-law form. The results are similar for the two methods.
Since our aim is to investigate the
effect of the core radius on strong gravitational lensing efficiency, we fit the
relation between $r_c$ and $M$ (Figure \ref{MRc}). As a check, we repeat the
procedure with independent data \citep{deblok02}. The results are
indistinguishable.

The gravitational lensing principle tells us that for any spherically symmetric
density profile (here, a CIS halo), multiple images of a source can be produced
only if the central convergence $\kappa_{c}$  is larger than
unity \citep{schneider92}. The central convergence is a measure of the central
surface mass density of the lensing halos. It is both mass and redshift
dependent. For singular density profiles such as SIS and NFW, the central value
is divergent, so $\kappa_{c}>1$ is always satisfied and multiple images can be
produced by any mass. For density profiles with a finite soft core, however, the
condition $\kappa_{c}>1$ imposes a minimum mass threshold to produce multiple
images. For CIS halos \citep{chen05}, we have $\kappa_c\propto M^{2/3}/r_c$. The
larger the core radius, the larger the mass needed to ensure $\kappa_{c}>1$.
While both the zero and maximum disk cases give similar $r_c$-$M$ relations, the
more conservative case is that with the smaller core radius for a given mass;
other choices would produce less lensing. We thus use the formula fit to the
zero disk case: $r_{c}=2.25(M/10^{12}M_\odot)^{1/3}$ kpc. Interestingly, this
formula is similar to the one derived analytically in the NTIS
model \citep{shapiro99,chen05}.

\begin{figure}
\includegraphics[width=8.5cm]{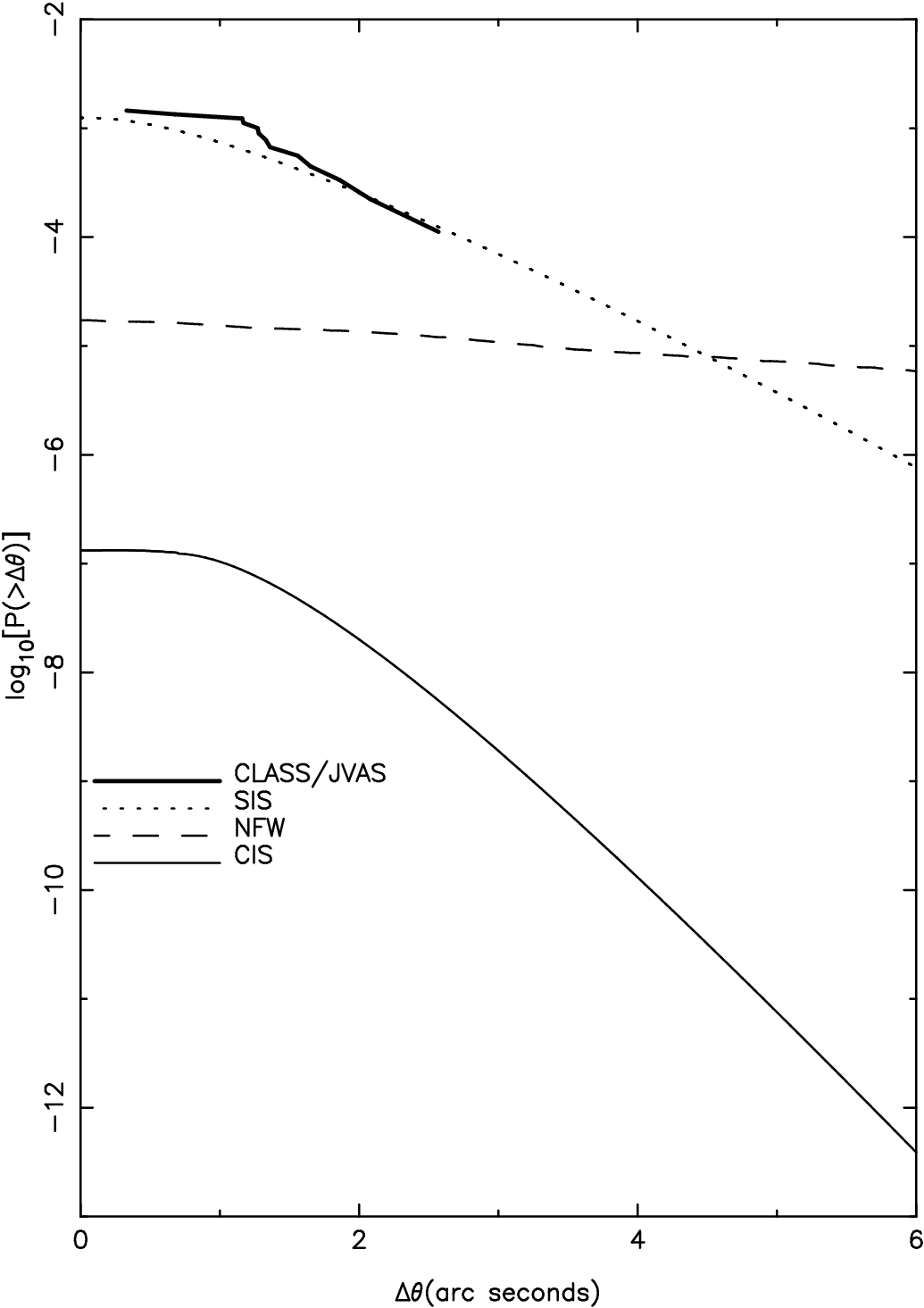}
\caption{The lensing probability with image separation larger than
$\Delta\theta$. Our prediction for the CIS model based on the observed $r_c$-$M$
relation (Figure \ref{MRc}) is shown as the solid line.  This fails to explain
the observed lensing frequency (heavy line) by four orders of magnitude.
In contrast, our approximate SIS model (dotted line) provides a reasonable
match to the data.  A pure NFW model (dashed line) gives intermediate
results.
\label{Ptheta}}
\end{figure}

The combined JVAS/CLASS survey forms a well-defined statistical sample
containing 13 multiply imaged lens systems \citep{myers03,browne03} among 8,958
sources.  These data provide the image separation $\Delta\theta$ for each lens
system.  The observational probability $P_{\mathrm{obs}}(>\Delta\theta)$
for the CLASS/JVAS survey is shown in Figure \ref{Ptheta}.

When a remote quasar is lensed by a CIS halo, three images are produced. The
image nearest the lens is very weak.  It disappears entirely when the source,
lens, and observer are aligned, and the Einstein ring appears. The image
separation $\Delta\theta$ is thus the separation between the outer two images.
By adopting a model for the density profile of lensing halos, their comoving
number density, and the geometry of the $\Lambda$CDM universe, we can predict
the properties of the strong lens systems.


In order to compare with the observed lensing probabilities, we calculate
$P_{\mbox{CIS}}(>\Delta\theta)$, the lensing probability for quasars at redshift
$z_s$ lensed by foreground CIS halos with image separation larger than
$\Delta\theta$. The redshift $z_s$ of the sources (quasars) for the CLASS/JVAS
sample has an approximately Gaussian distribution \citep{chae02,myers03} with a
mean of 1.27 and a dispersion of 0.95. The lensing rate is sensitive to $z_s$,
but the effect of the redshift distribution is negligible compared to the choice
of halo profile. We thus use the mean value $z_s=1.27$ in our calculations. For
each lens system, the image separation depends on the source position. For the
CIS model, however, the image separation is almost source position
independent \citep{chen05}, so we use the diameter of the Einstein ring as the
image separation for each lens system. Gravitational lensing magnifies the
brightness of sources, so the number of lenses will be
overrepresented \citep{turner84} in any observed sample. The theoretically
predicted lensing probability should therefore include a magnification bias (MB)
correction to the observed probability. The MB is calculated on the basis of the
total magnification of the outer two brighter images \citep{oguri02}.
One of the most important elements in predicting lensing probability is the
comoving number density of lensing galaxies. We adopt the results recently
derived \citep{choi07} from the Sloan Digital Sky Survey.
The background cosmology is taken from the five-year Wilkinson Microwave
Anisotropy Probe observations \citep{komatsu08}. The final predicted lensing
probability for CIS is plotted in Figure \ref{Ptheta}. For comparison, the
lensing probability of the SIS model is shown with the same parameters and
approximations as CIS. The NFW model \citep{chen05} is also shown.
This is important to modeling the lensing by clusters, but is not relevant
on the scale of individual galaxies considered here.

The predicted lensing probability for the CIS modeled 
dark matter halos is about four
orders of magnitude lower than the observations of CLASS/JVAS at all image
separations, and two orders of magnitude lower than the NFW model at small image
separations.
Though successful in fitting rotation curves, the CIS model is obviously
inadequate for explaining strong gravitational lensing.
We have used a spherical model. As it is known that the ellipticity
does not significantly affect the total lensing efficiency for SIS model\citep{huterer05} when 
compared with the inner density slope on galaxy scales. This is in contrast to 
galaxy clusters, in which the main inner density slope (NFW like, $\gamma\sim 1$)
 is shallower than SIS ($\gamma\sim 2$) and 
thus ellipticity and substructures would significantly increase 
the lensing efficiency\citep{bartelmann98,Meneghetti01,Meneghetti03,hennawi07,broadhurst08}.
Similarly, for large core size CIS model ($\gamma\sim 0$), lensing rate would become extremely
more sensitive to the lens shape and to external perturbations.
However, the combination of all our approximations together can shift the result
by no more than one order of magnitude, as can be seen from the close match of
our approximate SIS model to the data. So it is safe to conclude that dark halo
models like CIS and NTIS with the soft central cores derived from kinematic
observations can not account for the statistics of strong gravitational lensing
by themselves; they need a more centrally concentrated component like the
baryons.

It is not difficult to understand the low lensing probability of the CIS model.
Recall that the central convergence depends on the mass $M$ and the redshift
$z_L$ of the lensing halos. With the fitting formula $r_c\propto M^{1/3}$, we
have $\kappa_c(M,z_L)\sim M^{1/3}F(z_L)$, where
$F(z_L)=\Omega(z_L)^{1/6}D_LD_{LS}/D_S$, with
$\Omega(z_L)=\Omega_m(1+z_L)^3+\Omega_{\Lambda}$, $D_L$, $D_S$ and $D_{LS}$ are
the angular diameter distances from the observer to the lens, to the source and
from the lens to the source, respectively. For quasars at $z_S=1.27$, $F(z_L)$
has a maximum value of 0.24 for $z_L$ in the interval $[0,z_S]$.
The condition $\kappa_c>1$ for strong lensing implies $M>3\times
10^{13}M_\odot$. Since $M\sim 10^{13}M_\odot$ corresponds to the most massive
galaxies in the present universe, the galaxies with lower mass provide no
contribution to the total lensing probability. Furthermore, the contributions of
all galaxies to the total lensing probabilities are governed by the comoving
number density $n(M)$, which has a high-mass exponential cutoff \citep{chen08},
$n(M)\sim \exp(-M^{\beta/3})$, with $\beta=2.67$ in our calculations.
Consequently, galaxies with mass lower than $\sim 10^{13}M_\odot$ make no
contribution, and high-mass galaxies meet the exponential cutoff. Some
previous work \citep{hinshaw87,kochanek96,chiba99} also used the CIS model for early-type galaxies to
calculate the strong lensing probabilities, and obtained reasonable results.
They adopted a typical value of the core radius of $r_c\sim 0.1$ kpc, much smaller
than ours ($\sim 2.25$ kpc), so hardly different from SIS.

The only difference between CIS and SIS is that CIS has a finite core radius.
While the SIS model matches the lensing observations quite well, the low lensing
probabilities of the CIS model is in serious contradiction to
observations of strong gravitational lensing.  Similarly, the NFW/SIS model
contradicts rotation curve data. The proposed
remedy \citep{mcw06, mcw08} of the cusp/core problem via feedback driven
turbulence fixes this problem at the expense of creating another.

Most lensing galaxies are giant elliptical galaxies with substantial stellar
masses, while we base the CIS model on observations of dark matter dominated LSB
galaxies.  These are very different galaxy types.
Lensing is not sensitive to whether the mass doing the lensing is baryonic
or dark, so the contradiction might be avoided if the total mass distribution
of ellipticals --- stars plus dark matter --- can be modeled as SIS spheres.
The challenge then becomes a self-consistent understanding of the formation
of all galaxy types.

In the context of the $\Lambda$CDM structure formation paradigm,
the initial condition for galaxy formation is the NFW halo.  Baryonic
gas dissipates and settles to the center of the gravitational potential
defined by the dark matter to form the visible galaxy.  As the gas
collapses, the potential must adjust to the rearrangement of mass. This
process, commonly referred to as adiabatic
contraction \citep{barneswhite,gnedin04,sellwood05}, has the effect of
steepening \citep{dubinski} the mass profile (increasing $\gamma$).
Since the NFW halo is not adequate to explain lensing on its
own \citep{chen05,zhang04} (Figure \ref{Ptheta}), this process seems
necessary to produce elliptical galaxies that behave as SIS spheres.
Indeed, any transformation other than $\gamma = 1 \rightarrow 2$ would
fail to reproduce the lensing statistics.
However, this process cannot explain the observations of the rotation curves for
LSB galaxies (Figure \ref{u5750RC}).

In LSB galaxies, we need the opposite process: something that drives
$\gamma$ from $1 \rightarrow 0$.  This is what turbulence is
proposed \citep{mcw06, mcw08} to do.  The hypothesized turbulence
is driven by feedback from early star formation in the first halos.
If this process is universal and efficient, as proposed, then we may
only solve the cusp/core problem at the expense of introducing a
new problem with lensing.  The baryons must first collapse to the center
of the halo before they can drive feedback there.  So only one process can
dominate:  either adiabatic contraction, which increases $\gamma$,
or feedback, which might reduce $\gamma$.  If feedback succeeds in
establishing a soft core, it should persist through subsequent mergers
\citep{dehnen05, kzk06}.  It is difficult to see how an elliptical galaxy
with an SIS mass profile could be constructed in this scenario.

Nevertheless, this is what we need:  dark halos with a soft core that
persists in LSBs but elliptical galaxies that have a baryonic cusp.
Observationally, there is no clear objection to having elliptical galaxies
with a cuspy baryonic component embedded in a cored dark matter halo.
The problem comes in self-consistenly building both kinds of galaxies.

Dark matter can only interact with baryons through gravity.
The feedback of the baryons might re-shape the potential of the dark matter
and then the total mass distribution.
If strong outflows from stellar feedback carry dark matter particles
out of the central region via gravity, when baryons cool and collapse
to form the central baryonic cusp, they must necessarily 
bring back dark matter particles.  The non-adiabatic action of sudden
supernova driven outflows is only a minor perturbation on a zero sum
game \citep{gnedinzhao}.
In fact, recent simulations show that supernova-driven feedback 
inhibits the formation of baryonic
bulges (cuspy baryons) and decrease the dark matter density \citep{Governato2010}, so that the total
mass (baryons plus dark matter) density in the central regions of dwarf galaxies
would be core-like rather than cusp-like. 
If this process is generically effective at producing cores in dark matter
halos, then the early fragments that later build elliptical galaxies in 
$\Lambda$CDM should experience the same process.  Indeed, there is considerably
better evidence for strong star forming episodes elliptical
galaxies than in LSBs.  Once established, cores should persist
through subsequent mergers in the entire mass distribution, both dark
and baryonic.

We conclude that the apparent contradiction between rotation curves
and strong lensing statistics pointed out here is genuine.  It is difficult to
simultaneously reconcile the soft cored halos favored by many
kinematic observations with the singular mass profiles favored by
strong lensing.  In both cases, a fundamental tenet of the $\Lambda$CDM
structure formation paradigm, the NFW halo, is inadequate to explain
the observations.  Substantial rearrangement of the initial
NFW mass profile is required.  Ideas hypothesized to solve one
problem tend to make the other one worse.

\begin{acknowledgements}
This work was supported by the National Natural Science Foundation of China
under grant 11073023 and the National Basic
Research Program of China (973 Program) under grant 2009CB24901, and by the National
Science Foundation of the United States under grant AST0908370.
\end{acknowledgements}


\begin{thebibliography}{99}
\small \setlength{\itemindent}{-3mm} \setlength{\itemsep}{-0.5mm}
\setlength{\baselineskip}{4.5mm}

\bibitem[Barnes \& White(1984)]{barneswhite}Barnes, J., \& White, S. D. M. 1984, \mnras, 211, 753 
\bibitem[Bartelmann et al.(1998)]{bartelmann98}Bartelmann, M., et al. 1998, \aap, 330, 1
\bibitem[Begeman et al.(1991)]{begeman91}Begeman, K. G., Broeils,  A. H.,  \& Sanders, R. H. 1991,  \mnras, 249, 523
\bibitem[Broadhurst \& Barkana(2008)]{broadhurst08}Broadhurst, T. J., \& Barkana, R. 2008, \mnras, 390, 1647
\bibitem[Browne et al.(2003)]{browne03}Browne, I. W. A.,  et al. 2003, \mnras, 341, 13
\bibitem[Chae et al.(2002)]{chae02}Chae, K. -H., et al. 2002,  \prl, 89,
151301
\bibitem[Chen(2005)]{chen05}Chen, D. -M. 2005, \apj, 629, 23
\bibitem[Chen(2008)]{chen08}Chen, D.-M. 2008,    J. Cosmol. Astropart. Phys., 01, 006
\bibitem[Chiba \& Yoshii(1999)]{chiba99}Chiba, M.,  \& Yoshii, Y. 1999, \apj, 510, 42
\bibitem[Choi et al.(2007)]{choi07}Choi, Y. -Y.,  Park, C.,  \&  Vogeley, M. S. 2007,  \apj, 658, 884
\bibitem[Coles(2005)]{coles05}Coles, P. 2005,  Nature, 433, 248
\bibitem[de Blok \& Bosma(2002)]{deblok02}de Blok, W. J. G., \&Bosma,  A. 2002,   \aap  385, 816
\bibitem[de Blok \& McGaugh(1997)]{dBM97}de Blok, W. J. G.,  \& McGaugh, S. S. 1997,  \mnras, 290, 533
\bibitem[de Blok et al.(2001)]{dBMBR01}de Blok, W. J. G.,  McGaugh, S. S., Bosma, A.,  \&Rubin,  V. C. 2001, \apj,
552, L23
\bibitem[de Naray et al.(2008)]{denaray08}de Naray, R. Kuzio, McGaugh, S. S.,  \& de Blok, W. J. G. 2008,  \apj, 676, 920
\bibitem[de Naray et al.(2006)]{KdN06}de Naray, R. Kuzio, McGaugh,  S. S., de Blok, W. J. G.  \& Bosma, A. 2006, 
\apjs, 165, 461
\bibitem[Dehnen(2005)]{dehnen05} Dehnen, W. 2005,  \mnras, 360, 892
\bibitem[Dubinski(1994)]{dubinski}Dubinski, J. 1994, \apj,  431, 617
\bibitem[Gnedin et al.(2004)]{gnedin04}Gnedin, O.~Y., Kravtsov,  A.~V., Klypin,  A.~A.,  \& Nagai, D. 2004,  \apj, 616, 16
\bibitem[Gnedin \& Zhao(2002)]{gnedinzhao}Gnedin, O.~Y., \& Zhao, H. 2002,  \mnras, 333, 299
\bibitem[Governato et al.(2010)]{Governato2010}Governato, F.,  et al. 2010, Nature, 463, 203
\bibitem[Hennawi et al.(2007)]{hennawi07}Hennawi, J. F., Dalal, N., Bode, P., \& Ostriker, J. P. 2007, \apj, 654, 714
\bibitem[Hinshaw \& Krauss(1987)]{hinshaw87}
Hinshaw, G., \& Krauss, L. M. 1987, \apj, 320, 468
\bibitem[Huterer et al.(2005)]{huterer05}Huterer, D., Keeton, C. R.,  \& Ma, C.-P. 2005,  \apj, 624, 34
\bibitem[Jing(2000)]{jing00}Jing, Y. 2000,  \apj, 535, 30
\bibitem[Jing \& Suto(2002)]{jing02}Jing, Y., \& Suto, Y. 2002,  \apj, 574, 538
\bibitem[Kazantzidis et al.(2006)]{kzk06}Kazantzidis, S.,  Zentner, A. R.,   \&  Kravtsov, A. V. 2006,  \apj,  641, 647
\bibitem[Kochanek(1996)]{kochanek96}Kochanek, C. S. 1996,  \apj, 466, 638
\bibitem[Komatsu et al.(2009)]{komatsu08} E. Komatsu  et al. 2009, \apjs, 180, 330
\bibitem[Kormendy \& Freeman(2004)]{kormendy04}Kormendy, J.,  \& Freeman, K. C. 2004, Proceedings of IAU Symposium ``Dark Matter in Galaxies'', Eds. S. Ryder et al.,  220, 377
\bibitem[Li \& Chen(2009)]{li09}Li, N., \& Chen, D. -M. 2009, \raa, 9, 1173
\bibitem[Li \& Ostriker(2002)]{li02}Li, L. -X., \& Ostriker, J. P. 2002, \apj,  566, 652
\bibitem[Mashchenko et al.(2006)]{mcw06}Mashchenko, S., Couchman, H. M. P.,  \&Wadsley,  J. 2006,   Nature,  442, 539
\bibitem[Mashchenko et al.(2008)]{mcw08}Mashchenko, S., Wadsley,  J.,  \& Couchman, H. M. P. 2008    Science, 319, 174
\bibitem[McGaugh(2007)]{mcgaugh07}McGaugh, S. S.  et al. 2007,  \apj, 659, 149
\bibitem[McGaugh et al.(2001)]{MRB01}McGaugh, S. S., Rubin, V.C.,  \& de Blok, W. J. G. 2001, 
\aj, 122, 2381
\bibitem[Meneghetti et al. (2001)]{Meneghetti01}Meneghetti, M. et al. 2001, \mnras, 325, 435
\bibitem[Meneghetti et al. (2003)]{Meneghetti03}Meneghetti, M., Bartelmann, M, \& Moscardini, L. 2003, \mnras, 340, 105
\bibitem[Myers et al.(2003)]{myers03}Myers, S. T.,  et al. 2003, \mnras, 341, 1
\bibitem[Navarro, Frenk \& White(1997)]{nfw97}Navarro, J. F., Frenk,  C. S., \& White, S. D. M. 1997,   \apj, 490,
493
\bibitem[Navarro et al.(2004)]{nfw04}Navarro, J. F.,  et al. 2004, \mnras, 349, 1039
\bibitem[Romanowsky et al.(2003)]{roman03}Romanowsky, A. J.,  et al. 2003, Science, 301, 1696
\bibitem[Oguri et al.(2002)]{oguri02}Oguri, M., Taruya,  A., Suto, Y.,  \& Turner, E. L. 2002,   \apj,
568, 488
\bibitem[Oguri et al.(2008)]{oguri08}Oguri, M.,   et al. 2008, \aj, 135, 512
\bibitem[Salucci et al.(2007)]{salucci07}Salucci, P.,  et al. 2007, \mnras,  378, 41
\bibitem[Schneider et al.(1992)]{schneider92}Schneider, P.,  Ehlers, J.,  \& Falco, E. E. 1992,  {\it
Gravitational lenses}(Springer-Verlag, Berlin), p.230
\bibitem[Sellwood \& McGaugh(2005)]{sellwood05}Sellwood,  J. A.,  \& McGaugh, S. S. 2005,   \apj, 634, 70
\bibitem[Shapiro et al.(1999)]{shapiro99}Shapiro, P. R., Iliew, I. T.,  \& Raga,   A. C. 1999,   \mnras,
307, 203
\bibitem[Turner et al.(1984)]{turner84}Turner, E. L., Ostriker,  J. P., \& Gott, J. R. 1984,   \apj, 284,
1
\bibitem[van der Hulst et al.(1993)]{vdHulst93}van der Hulst, J. M.  et al. 1993, \aj, 106, 548
\bibitem[Zhang(2004)]{zhang04}Zhang, T.-J. 2004,  \apj, 602, L5
\end{thebibliography}
\end{document}